# Screening Promising Thermoelectric Materials in Binary Chalcogenides through High-Throughput Computations

Tiantian Jia[1,2], Zhenzhen Feng[1,2], Shuping Guo[1,2], Xuemei Zhang[1,2] and Yongsheng Zhang[1,2*]

1) Key laboratory of Materials Physics, Institute of Solid State Physics, Chinese Academy of Sciences, 230031 Hefei, P. R. China

2) Science Island Branch of Graduate School, University of Science and Technology of China, 230026 Hefei, P. R. China

Corresponding author:
yshzhang@theory.issp.ac.cn (Y. Zhang)

## Abstract

The high-throughput (HT) computational method is a useful tool to screen high performance functional materials. In this work, using the deformation potential method under the single band model, we evaluate the carrier relaxation time and establish an electrical descriptor ($\chi$) characterized by the carrier effective masses based on the simple rigid band approximation. The descriptor ($\chi$) can be used to reasonably represent the maximum power factor without solving the electron Boltzmann transport equation. Additionally, the Grüneisen parameter ($\gamma$), a descriptor of the lattice anharmonicity and lattice thermal conductivity, is efficiently evaluated using the elastic properties, omitting the costly phonon calculations. Applying two descriptors ($\chi$ and $\gamma$) to binary chalcogenides, we HT compute 243 semiconductors and screen 50 promising thermoelectric materials. For these theoretically determined compounds, we successfully predict some previously experimentally and theoretically investigated promising thermoelectric materials. Additionally, 9 *p-type* and 14 *n-type* previously unreported binary chalcogenides are also predicted as promising thermoelectric materials. Our work provides not only new thermoelectric candidates with perfect crystalline structure for the future investigations, but also reliable descriptors to HT screen high performance thermoelectric materials.

# I.   Introduction

Thermoelectric materials can directly convert heat into electricity. They play many promising roles in solving problems of energy and environmental crisis. However, the conversion efficiencies of current discovered thermoelectric materials are about 5-20%[1], limiting the global commercialization. The conversion efficiency is characterized by the thermoelectric figure of merit ($ZT$), $ZT = S^2\sigma/\kappa$, where $S$ is the Seebeck coefficient, $\sigma$ is the electrical conductivity and $\kappa$ is the thermal conductivity [coming from two sources: the carrier thermal conductivity ($\kappa_e$) and the dominating lattice thermal conductivity ($\kappa_l$)]. Therefore, high performance thermoelectric materials must possess both excellent electrical transport properties to enhance $S$ and σ, and strong phonon scatterings to decrease $\kappa_l$.

According to the electron Boltzmann transport theory[2], $S$ and σ can be determined by the electronic properties of materials, but they are entangled with each other through the carrier concentration. It is difficult to enhance $S$ and σ simultaneously. In recent years, many methodologies have been developed to increase the power factor ($PF = S^2\sigma$). The band structure engineering is one of the most important methods to optimize the $PF$, such as enhancing the Seebeck coefficient through generating resonant states, multiple-band convergence mechanism or the energy filtering; and raising the electrical conductivity through increasing the carrier concentration or preserving the carrier mobility, etc.[3-7] Additionally, $\kappa_l$ is an important thermoelectric parameter, strongly depending on phonon scatterings. The nanostructure engineering (introducing the atomic-scale disorder or forming nanoscale endotaxial precipitates, mesoscale grain boundaries, nanocomposites, and nanostructurings, etc) [6-9] is used to decrease the thermal conductivity by producing the multiscale phonon scattering regions, but which can also potentially decrease the electrical conductivity by scattering electrons as well.

Alternatively, in fact, some compounds with ordered crystal structures exhibit intrinsically low thermal conductivity due to strong lattice anharmonicity[10,11], and do not affect the electrical conductivity.[12] In 2014, Zhao et al.[13] reported that SnSe exhibits an unprecedented maximum $ZT$ value (~2.6) at 923 K due to the ultralow lattice thermal conductivity induced by the strong anharmonic bonding. Therefore, searching for crystalline compounds with strong anharmonicity and good electrical properties is one of the most important topics in the thermoelectric community.

The state-of-the-art commercial thermoelectric materials with relatively high *ZT* values are limited to few semiconductors (such as $Bi_2Te_3$, PbTe, SiGe, etc.), in which the binary chalcogenides ($Bi_2Te_3$ and PbTe) based thermoelectric materials are extensively-studied.[14-16] In addition, many compounds from the binary chalcogenide family are traditionally suggested as promising thermoelectric materials, such as PbX (X=S, Se)[16,17], $Cu_2Se$[18,19], $In_4Se_3$[20], $La_3Te_4$[21], etc[22-27]. Moreover, from time to time, some new high performance binary chalcogenides are discovered experimentally or predicted theoretically, such as the above mentioned SnSe compound[13] and a theoretically predicted GeSe compound[28] (an excellent thermoelectric material due to the strong anharmonicity and the multiband convergence effect). All these studies indicate that many compounds from the binary chalcogenide family might have intrinsically low thermal conductivity and good electrical transport properties, which can be considered as promising thermoelectric materials. Based on the Crystallography Open Database (COD)[29-33], there are 668 binary chalcogenides and the thermoelectric properties of most of them have not yet been examined by either experimental measurements or theoretical calculations. Therefore, we are going to use the high-throughput (HT) computational method to screen promising thermoelectric materials within these binary chalcogenides.

The HT computational method is a useful tool to screen high performance functional materials. Due to the complexity of syntheses and characterizations, the detailed experimental evaluation of the functional properties is usually time-consuming and costly. Therefore, it is difficult to find high performance functional materials through experimental screening from a large amount of compounds. Density functional theory (DFT) calculations are typically quite accurate for evaluating the properties of functional materials and can be easily performed using supercomputers. In recent years, DFT based HT theoretical calculation approach offers a more efficient and less costly method to discover new functional materials[34-50], such as lithium-ion batteries[35], catalysts[36] and thermoelectric materials[37-51], etc. However, there are two limitations in many previous thermoelectric HT computations: (1) calculating the carrier relaxation time ($\tau$) for the electrical transport property ($\sigma$); (2) computing the phonon property for the lattice thermal conductivity ($\kappa_l$).

For good electrical conductors, $\tau$ is almost energy-independent[52], allowing the constant relaxation time approximation (CRTA, $\tau$ can be set as a constant for different electron energy) were usually used in HT electronic transport calculations.[37-39] In 2008, Yang et al.[38] used the CRTA to HT evaluate the thermoelectric-related electronic transport properties of over 100 half-Heusler alloys and successfully predicted several new half-Heusler compounds with high power factors. Again, based on the simple CRTA (by assuming $\tau = 10^{-14}$ s for different compounds at different temperatures and carrier concentrations), Gibbs et al.[39] investigated the relationship between the maximum power factor and the Fermi surface complexity factor for ~2300 cubic compounds. They suggested that the Fermi surface complexity factor is an effective descriptor for the electrical properties. From these above HT studies, some new promising thermoelectric materials have been predicted. However, Comparing the computed thermoelectric properties of more than 48000 inorganic compounds with experimental data, Chen et al.[40] suggested that the CRTA gives rise to a high inaccuracy in the electrical conductivity calculations. Therefore, recently, in order to improve the prediction ability in evaluating $\sigma$, instead of using the CRTA, the high accuracy computations were used in HT computations[41,42], such as Xi et al.[41] used the deformation potential approximation method[53] to evaluate the carrier relaxation times and electrical properties of the chalcogenides with diamond-like structures, and successfully predicted a new series of diamond-like ternary chalcogenides possessing relatively higher electrical thermoelectric properties. Georgy et al.[42] proposed the electron-phonon averaged (EPA) approximation approach and deployed the EPA to HT screen the promising thermoelectric compounds from the wide half-Heusler family.

To evaluate $\kappa_l$ in HT computations, there were several different methods been proposed in previous researches.[43-47] Due to the strong phonon scattering phenomenon of the lone-pair electrons, Nielsen, et al.[43] used the minimum (the amorphous limit) thermal conductivity formulas to HT screen the low thermal conductivity materials from 72 compounds with lone-pair electrons. Through assuming that Grüneisen parameter is material independent, Gorai et al.[44] fitted the empirical thermal conductivity parameters using the experimental thermal conductivity of 45 compounds. And by combining with the empirical carrier mobility parameters fitted using the experimental carrier mobility of 31 compounds, they then came up with a semi-empirical descriptor to evaluate the thermoelectric properties of 518 $A_1B_1$ compounds.[45] Afterwards, to improve the accuracy of their semi-empirical model, Gorai et al.[46] re-fitted these empirical parameters by explicitly treating the van der Waals interactions in quasi-2D materials, and successfully predicted some known and new quasi-2D thermoelectric materials. Moreover, to consider the impact of anharmonicity on $\kappa_l$, Miller et al. [47] further integrated the coordination number in the semi-empirical thermal conductivity descriptor and improved the evaluation accuracy of $\kappa_l$ in HT thermoelectric computations. Additionally, by calculating the Debye temperature and Grüneisen parameter, Slack equation methods were proposed for HT lattice thermal conductivity calculations.[48-51] Such as, Curtarolo et al.[48-50] successfully incorporated the "GIBBS" quasiharmonic Debye mode in Automatic GIBBS Library (AGL), and used it to calculate and verify $\kappa_l$ of 75 materials with the diamond, zinc-blende, rocksalt, and wurzite structures, and 107 half-Heusler compounds. Recently, we also established a methodology to determine $\kappa_l$ using computationally feasible elastic properties (the bulk and shear modulus).[51]

In this work, by solving the electron Boltzmann transport equation (BTE)[2] under the CRTA, we firstly calculate partial electron transport properties ($S$ and $\sigma/\tau$). To improve the accuracy in the evaluation of σ, we calculate τ by using the deformation potential method under the single band model[53]. And then establish and validate an electrical descriptor ($\chi$) to characterize the maximum thermoelectric power factor. For the lattice thermal conductivity of a perfect ordered compound, it is characterized by the lattice anharmonicity or the Grüneisen parameter ($\gamma$), which are efficiently evaluated using the elastic properties[51]. Based on the two well defined descriptors ($\chi$ and $\gamma$), we investigate the thermoelectric properties of 243 known binary semiconductor chalcogenides ($Y_nX_m$, where Y is a cation, X is an anion of S, Se or Te, and n and m are integers), and screen 50 high performance thermoelectric materials. Our HT calculations successfully predict some previously experimentally and theoretically studied thermoelectric compounds. Additionally, 9 and 14 previously unreported binary chalcogenides are predicted as promising *p-type* and *n-type* thermoelectric materials, respectively.

## II. Methodologies

### A. Electrical property calculations

Electrical transport properties (the Seebeck coefficient and the electrical conductivity) play a vital role in determining high performance thermoelectric materials. To calculate the properties, we should know the carrier concentration and its distribution responding to temperature and chemical potential. The electron BTE[2] is a powerful tool to solve the non-equilibrium probability distribution of carriers. Based on the relaxation time approximation, the Boltzmann transport theory has been used to calculate the electrical transport properties using electronic band structures.[54] According to the electron BTE, the electrical conductivity tensors ($\sigma_{\alpha\beta}$) and Seebeck coefficient tensors ($S_{\alpha\beta}$) can be written as a function of temperature ($T$) and chemical potential ($\mu$):

$$\sigma_{\alpha\beta}(E) = e^2 \tau g(E) v_\alpha(E) v_\beta(E) \quad (1)$$

$$\sigma_{\alpha\beta}(T;\mu) = \int \sigma_{\alpha\beta}(E) \left[ -\frac{\partial f_0(T;E;\mu)}{\partial E} \right] dE \quad (2)$$

$$S_{\alpha\beta}(T;\mu) = -\frac{\frac{1}{eT}\int \sigma_{\alpha\beta}(E)(E-\mu)\left[-\frac{\partial f_0(T;E;\mu)}{\partial E}\right]dE}{\int \sigma_{\alpha\beta}(E)\left[-\frac{\partial f_0(T;E;\mu)}{\partial E}\right]dE} \quad (3)$$

where $\alpha$ and $\beta$ ($\alpha, \beta = x, y, z$) are the tensor indices, $\tau$ is the relaxation time, $f_0$ is the Fermi–Dirac distribution function, $g(E)$ is the density of states, and $v_\alpha(E)$ is the group velocity. $g(E)$ and $v_\alpha(E)$ can be directly derived from the electronic band structures. Under the simple rigid band approximation (RB), the carrier concentration $n$ can be written as[55]

$$n(T;\mu) = \int_{-\infty}^{E_V} g(E)[1 - f_0(T;E;\mu)]dE - \int_{E_C}^{\infty} g(E) f_0(T;E;\mu) dE \quad (4)$$

where $E_V$ and $E_C$ are the energy of the valence band maximum (VBM) and conduction band minimum (CBM), respectively. Since the promising thermoelectric materials are typically heavily doped semiconductors with a carrier concentration between $10^{19}$ and $10^{21}$ cm$^{-3}$[56], we focus on the theoretically predicted thermoelectric properties within this concertation range and temperature is considered from 200 K to 1000 K at intervals of 50 K.

In order to investigate the electronic transport properties (Eq. 1-3), the carrier relaxation time ($\tau$) must be given. In principle, $\tau$ is determined by several scattering mechanisms, such as the electron–phonon scattering, the impurity scattering, etc. For different scattering mechanisms, $\tau$ can be written as[55]

$$\tau = \tau_0 \frac{2}{3}\left(r + \frac{3}{2}\right)\frac{F_{r+\frac{1}{2}}(\eta)}{F_{\frac{1}{2}}(\eta)} \tag{5}$$

where $F_x(\eta) = \int_0^\infty \frac{E^x}{1+\exp(E-\eta)} dE$, $\eta = \frac{\mu}{k_B T}$, $\eta$ is the reduced chemical potential, $k_B$ is the Boltzmann constant, $\tau_0$ is independent of energy, and $r$ is the scattering parameter ($r = -1/2$ for the acoustic phonon scattering, $r = 3/2$ for the ionized impurity scattering, and $r = 1/2$ for the polar optical phonon scattering.). Obviously, $\tau$ strongly depends on the temperature and chemical potential. Therefore, in this work, we are committed to calculate the $\tau$ at different temperatures and carrier concentrations.

In general, the acoustic phonon scattering is the dominant carrier scattering mechanism.[53] Moreover, considering the approximations of single band and spherical Fermi surface, $\tau$ can be systematically determined from the first-principles calculations by explicitly considering the acoustic phonon scattering mechanisms. In the model, $\tau$ is calculated by the deformation potential method as[53]

$$\tau(T;\mu) = \frac{2^{\frac{1}{2}}\pi\hbar^4 \rho v_l^2}{3E_d^2(m_c^* k_B T)^{\frac{3}{2}}} \frac{F_0(\eta)}{F_{\frac{1}{2}}(\eta)} \tag{6}$$

where $\hbar$ is the reduced Planck constant, $\rho$ is the mass density, $v_l$ is the longitudinal sound velocity, $m_c^*$ is the conductivity effective mass, and $E_d$ is the deformation potential (strength of the electron-acoustic interaction). And $v_l$ can be calculated from the elastic properties [bulk modulus (B) and shear modulus (G)], as $v_l = \sqrt{\frac{B+4/3G}{\rho}}$;[57,58] $m_c^*$ can be computed based on the band structures; $E_d$ is defined as $E_d = \Delta E/(\frac{\Delta V}{V})$, where $\Delta E$ is energy change of the band extrema with the volume dilation ($\frac{\Delta V}{V}$). Sometime, this theoretically calculated $E_d$ might not well agree with the experimental measurements by fitting the carrier mobility. The discrepancy of $E_d$ between the theoretical calculations and the experimental measurements might be due to the following reasons: (1) experimental results depend on the chosen transport mechanism, and the corresponding parameters are fitted using a relatively simple model; (2) the actual carrier mobility includes the effects of impurity scattering, polar optical scattering and acoustic phonon scattering, etc. Obviously, the impurity and optical scatterings are not considered in the current theoretical calculations; (3) the experimental measurements are largely diverged, e.g. 20-60 eV for *p-type* PbTe[59-61]. Luckily, the theoretically calculated $E_d$ can agree with each other very well, e.g. *p-type* PbTe (12.3 eV from our work and 10.5 eV from Ref. 58) and *p-type* SnSe (11.5 eV from our work and 15.6 eV from Ref. 60). Nevertheless, although the current method might not be perfect, many relevant literatures studying the chalcogendies (such as SnSe$_2$[23], SnX[62], and CuBiS$_2$[63], etc.[64,65]) have used it to evaluate $E_d$ and $\tau$ due to its efficiency. The HT computations do require screening functional materials efficiently (fast with reasonable accuracy). This method is a suitable trade-off methodology in the HT computations. Very recently, Xi et al.[42] also used the method to calculate the deformation potentials of the diamond-like chalcogenides in HT computations. Additionally, to check the reasonableness of the deformation potential method, we calculate the carrier relaxation times ($\tau_{DP}$, Fig. S1a) and corresponding electrical conductivities ($\sigma_{DP}$, Fig. S1b) of *p-type* PbSe (225) at $1.5 \times 10^{20}$ cm$^{-3}$ at different temperatures. It turns out that the calculated $\sigma_{DP}$ are in reasonable agreement with the experiment measurements ($\sigma_e$), and $\sigma_{DP}$ and $\sigma_e$ show the same tendency as the temperature changes. However, under the simple CRTA (the carrier relaxation time to be a constant (10$^{-14}$ s) at different temperatures), the calculated electrical conductivities ($\sigma_{CRTA}$) are significantly different to $\sigma_e$. This means that the deformation potential method can be used to reasonably evaluate the carrier relaxation time of a compound, and we are going to use this method in our HT computations.

**B. An electrical property descriptor: $\chi$**

To obtain the electrical transport properties by solving the electron BTE, the Brillouin zone should be sampled by a highly dense k-point mesh, which is computationally cost and is not suitable for HT computations. To screen high performance thermoelectric materials efficiently, we require some computationally addressable metrics (or descriptors) that can reasonably quantify the electrical properties. Herein, we are going to propose an electrical descriptor to characterize the thermoelectric power factor ($PF$).

For a heavily doped semiconductor, the Seebeck coefficient ($S$) can be simply evaluated at different temperatures ($T$) and carrier concentrations ($n$) as[55]

$$S = \frac{2k_B^2 T}{3e\hbar^2} m_d^* \left(\frac{\pi}{3n}\right)^{\frac{2}{3}} \qquad (7)$$

where $m_d^*$ is the density of states (DOS) effective mass. The electrical conductivity ($\sigma$) can be evaluated from[55]

$$\sigma = ne\mu_e = \frac{ne^2}{m_c^*}\tau \qquad (8)$$

where $\mu_e$ is the carrier mobility. Using the Fermi integrals, $n$ can be expressed as[55]

$$n = \frac{1}{2\pi^2}\left(\frac{2m_d^* k_B T}{\hbar^2}\right)^{\frac{3}{2}} F_{\frac{1}{2}}(\eta) \qquad (9)$$

Combining with the equation of $\tau$ (Eq. 6), the power factor ($PF$) can be rewritten as

$$PF = S^2 \sigma = \left[\frac{2k_B^2 T}{3e\hbar^2} m_d^* \left(\frac{\pi}{3n}\right)^{\frac{2}{3}}\right]^2 \frac{ne^2}{m_c^*} \frac{2^{\frac{1}{2}}\pi\hbar^4 \rho v_l^2}{3E_d^2(m_d^* k_B T)^{\frac{3}{2}}} \frac{F_0(\eta)}{F_{\frac{1}{2}}(\eta)} \propto \frac{(m_d^*)^{3/2}}{(m_c^*)^{5/2}} \frac{\hbar k_B^2 \rho v_l^2}{E_d^2} \frac{F_0(\eta)}{\left[F_{\frac{1}{2}}(\eta)\right]^{\frac{4}{3}}} \qquad (10)$$

where, $\dfrac{F_0(\eta)}{\left[F_{\frac{1}{2}}(\eta)\right]^{\frac{4}{3}}}$ is the function of external environment parameters ($T$ and $\mu$). We then define the rest part as a thermoelectric parameter ($\chi$)

$$\chi = \frac{(m_d^*)^{3/2}}{(m_c^*)^{5/2}} \frac{\hbar k_B^2 \rho v_l^2}{E_d^2} \qquad (11)$$

From Eq. 11, we realize that $\chi$ has the same dimension as $PF$. Thus, $\chi$ can be used as an electrical descriptor to simply characterize $PF$.

To evaluate $\chi$, we need to efficiently calculate $m_c^*$ and $m_d^*$. Since the realistic $m_d^*$ and $m_c^*$ depends on temperature, doping, etc., it is pretty hard to accurately evaluate them. Gibbs et al.[39] suggested that $m_d^*$ and $m_c^*$ can be evaluated from the calculated $S$ and $\sigma$ by solving the electron BTE, representing as $m_d^{BT^*}$ and $m_c^{BT^*}$

$$m_d^{BT^*}(T;\mu) = \frac{\hbar^2}{2k_B T}\left[\frac{2\pi^2 n(T;\mu)}{F_{1/2}(\eta)}\right]^{\frac{2}{3}} \qquad (12)$$

$$\left[m_c^{BT^*}(T;\mu)\right]^{-1} = \frac{\sigma(T;\mu)}{e^2 \tau} \times \frac{1}{n(T;\mu)} \qquad (13)$$

Additionally, since most promising thermoelectric materials possess multiple energy valleys and non-spherical Fermi surfaces, Gibbs et al.[39] have defined a Fermi surface complexity factor ($N_V K^*$) as

$$N_V K^* = \left(\frac{m_d^*}{m_c^*}\right)^{3/2} \quad (14)$$

where $N_V$ is the valley degeneracy, $K^*$ is the anisotropy parameter as

$$N_V = \left(\frac{m_d^*}{m_b^*}\right)^{3/2}, K^* = \left(\frac{m_b^*}{m_c^*}\right)^{3/2} \quad (15)$$

where $m_b^*$ the single valley band effective mass. Additionally, under the CRTA, Gibbs et al.[39] also found that the $N_V K^*$ has a nearly linear relationship with the electron BTE calculated maximum $PF$ ($PF_{\max}$). In this work, to eliminate the limitations of CRTA, through considering the acoustic phonon scattering effect in $\tau$ (Eq. 6), we are committed use $\chi$ instead of $N_V K^*$ to evaluate $PF$ (Eq. 10). Obviously, we can use the BTE calculated effective masses ($m_d^{BT^*}$ and $m_c^{BT^*}$) to calculate $\chi$ ($\chi^{BT} = \frac{\left(m_d^{BT^*}\right)^{3/2}}{\left(m_c^{BT^*}\right)^{5/2}} \frac{\hbar k_B^2 \rho v_l^2}{E_d^2}$).

However, to avoid solving the electron BTE in future HT calculations, we are committed to straightforwardly evaluate effective masses based on the rigid band (RB) approximation. According to the RB approximation, the relationship between the density of states [$g(E)$] and the electronic energy ($E$) is

$$g(E) = \frac{1}{2\pi^2}\left(\frac{2m_d^{RB^*}}{\hbar^2}\right)^{\frac{3}{2}} E^{\frac{1}{2}} \quad (16)$$

And the RB determined hole and electron DOS effective mass ($m_d^{RB^*}$) can be approximately calculated by fitting the relationship between $g(E)$ and $E$ around the VBM and CBM, respectively. Additionally, based on the equipartition principle[55], the RB determined conductivity effective mass ($m_c^{RB^*}$) can be evaluated from the harmonic mean effective mass of single energy valley, as

$$\frac{1}{m_c^*} = \frac{1}{3}\left(\frac{1}{m_x} + \frac{1}{m_y} + \frac{1}{m_z}\right), m_\alpha = \hbar^2 \left[\frac{\partial^2 E(k)}{\partial_\alpha^2}\right]^{-1} \quad (\alpha = x, y, z) \quad (17)$$

where $m_\alpha$ is the band effective mass along the $\vec{\alpha}$ direction. Therefore, through the efficiently calculating $m_d^{RB^*}$ and $m_c^{RB^*}$ using the DOS and band structures, we can obtain the RB determined $\chi$ ($\chi^{RB} = \frac{\left(m_d^{RB^*}\right)^{3/2}}{\left(m_c^{RB^*}\right)^{5/2}} \frac{\hbar k_B^2 \rho v_l^2}{E_d^2}$) as well, which will be a feasible and suitable metric for evaluating the electrical transport properties (such as $PF$) of thermoelectric materials.

## C. A lattice anharmonic descriptor: $\gamma$

After establishing an electrical descriptor ($\chi$), we turn to the other important thermoelectric parameter, the lattice thermal conductivity ($\kappa_l$). There are many different lattice thermal conductivity formulas, such as the Slack model, the Callaway model, the Debey-Callaway model, and etc. These formulas usually give rise to quite different lattice thermal conductivity values. Although there are several different forms for lattice thermal conductivity that differ in detail, all of them show that the Grüneisen parameter ($\gamma$) is an important parameter to evaluate $\kappa_l$. In this work, we'd rather use $\gamma$ as the descriptor, instead of a "real" $\kappa_l$. $\gamma$ represents the strength of lattice anharmonicity of a compound, and $\kappa_l$ is proportional to $1/\gamma^2$.[66] However, to accurately obtain $\gamma$, the qusi-harmonic phonon dispersions[67] or the time-consuming second- and third-order force constants required phonons[68,69] should be calculated. Recently, we have developed an effective method to evaluate $\gamma$ and $\kappa_l$ through the computationally feasible elastic properties[51]. The methodology has been validated using ~40 different compounds from three different prototype structures (rocksalt, zinc blende and wurtzite) and the well-studied SnSe compound. It turns out that the calculated $\gamma$ through using the elastic properties are in reasonable agreement with those using the quasi-harmonic phonon calculations. And the elastic properties evaluated $\kappa_e$ (or the experimentally measured $\kappa_{exp}$) is about 2.5 times of $\kappa_{exp}$ (or $\kappa_e$), $\frac{1}{40}\sum\frac{\max(\kappa_e,\kappa_{exp})}{\min(\kappa_e,\kappa_{exp})} \approx 2.5$, which is comparable with those from semi-empirical model[44]. In the elastic evaluation method, $\gamma$ is calculated from the elastic properties (B and G), as[51]

$$\gamma = \sqrt{\frac{(\gamma_l)^2 + 2(\gamma_S)^2}{3}}, \quad \gamma_l = -\frac{1}{2}\frac{V}{B+\frac{4}{3}G}\frac{\partial\left(B+\frac{4}{3}G\right)}{\partial V} - \frac{1}{6}, \quad \gamma_S = -\frac{1}{2}\frac{V}{G}\frac{\partial G}{\partial V} - \frac{1}{6} \qquad (18)$$

where $\gamma_l$ and $\gamma_S$ are the longitude and shear Grüneisen parameter, and V is the volume of unit cell. From Eq. 18, we find that $\gamma$ characterizes the relationship between the elastic properties and V change. The method provides a computationally cheap but reasonable way to consider the anharmonicity of a compound. Therefore, in this work, the $\gamma$ of all binary semiconductor chalcogenides are calculated using the efficient elastic evaluation method, and used as the anharmonicity descriptor to screen compounds with strong lattice anharmonicity or low lattice thermal conductivity.

## D. Density-functional theory calculations

The DFT calculations are performed in the Vienna Ab initio Simulation Package (VASP).[70] The interactions between electrons are described by the projector augmented-wave (PAW) method.[71] The energy cutoff for the plane-wave expansion is set within 262 - 600 eV according to elements in different compounds. The electronic exchange-correlation functional is accounted for by the generalized gradient approximation of Perdew, Burke, and Ernzerhof (GGA-PBE).[72] The geometry is relaxed until the total energy is less than $10^{-5}$ eV and the forces are below 0.01 eV/Å. The Brillouin zones are sampled by Monkhorst-Pack $k$-point meshes[73] for all compounds with meshes chosen to give a constant density of $k$ points (30 Å$^3$). To obtain accurate electrical transport properties ($\sigma$ and $S$), more dense k-mesh (100 Å$^3$ k-point density) is used in solving BTE. We perform the BoltzTraP code[54] to solve the BTE under the relaxation time and rigid band approximations. Additionally, it is worth pointing out that despite some binary chalcogenides include transition metal or rare-earth elements, due to the correct Hubbard U values or different magnetic configurations (especially nonlinear magnetic configurations) are difficult treated in the HT work, we just carry out the normal DFT-PBE calculations in our HT computations[38,41].

The accurate geometry is important to obtain reasonable elastic properties for evaluating $\gamma$. Since the PBEsol functional[74] is suggested to give rise to accurate geometric results, such as lattice constants,[75] it is used to evaluate the elastic properties of compounds. Additionally, the elastic constants ($C$ or $c_{ij}$) are calculated from the strain-stress relationship[76]: $\sigma_s = C \cdot \epsilon$, where $\sigma_s$, $C$, and $\epsilon$ are the engineering stress vector, the stiffness matrix, and the strain vectors, respectively. According to the Voigt-Reuss-Hill (VRH) theory[77] in a macroscopic system, the corresponding elastic properties, such as B and G, can be evaluated from the elastic constants. Additionally, in the calculations of $\gamma$ and $E_d$, the crystal volumes are isotropically changed ~+/-2%. Considering the energy of the deep core level is hardly affected by such small volume variations, we choose the deep 1s level as the reference level in the $E_d$ calculations.

We write a thermoelectric HT computation code, named ThermoElectric Material Genome (TEMG), which can be used to automatically setup HT computations, including converting the geometric structures of compounds in the database into the VASP input structure files, preparing VASP-DFT input parameters, submitting DFT calculation jobs, collecting DFT calculated results and analyzing the (thermoelectric) properties of materials.

## III.  Semiconductor screening

From the COD database, there are 741 binary chalcogenides. We remove the compounds with partially occupied atoms, which are not suitable for the DFT calculations. Additionally, to release the computational burden, we intend to remove superlarge systems (containing more than 100 atoms and the cell volume larger than 2000 Å$^3$). This leads to 668 binary chalcogenide compounds in our computational database. We calculate the band structures of these compounds (the stoichiometry chemical formulas and the corresponding space group numbers of these compounds are included in Supplementary Tables). Based on the calculated band gap ($E_g$), we find that 360 binary chalcogenides are metals (with a zero band gap) and the rest 308 compounds are nonmetals (semiconductors and insulators shown in Supplementary Tables). Although there has the well-known band gap underestimation using the PBE functional, the previous HT results based on PBE are still useful in predicting promising thermoelectric materials.[38,42]

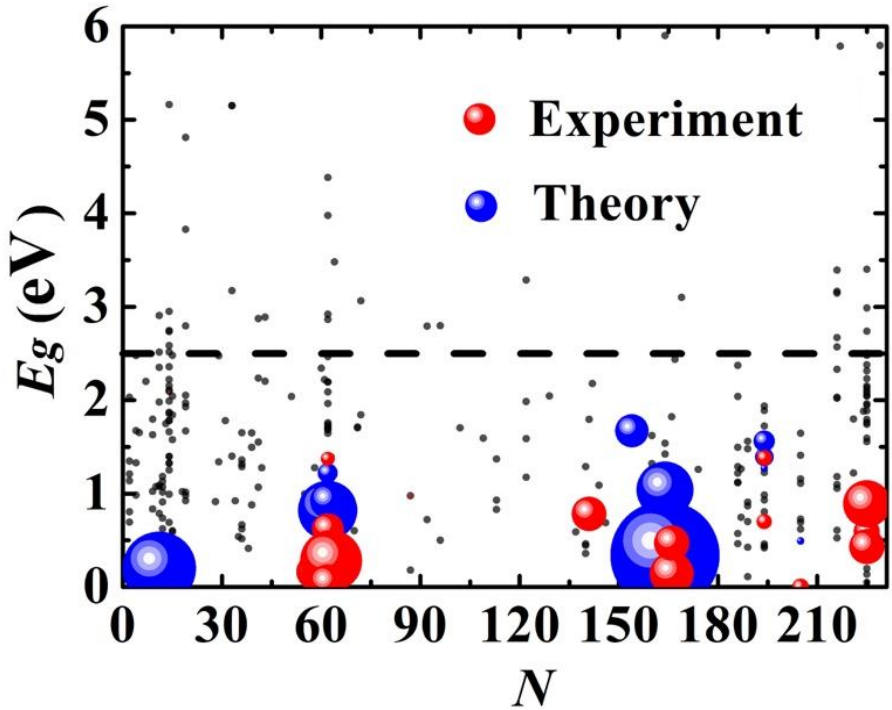

Figure 1. Theoretically calculated band gaps ($E_g$) of 308 nonmetal binary chalcogenides versus their space group numbers ($N$). The blue and red circles represent the previously experimentally and theoretically investigated promising binary chalcogenides, respectively. The size of the circle represents the peak $ZT$ value of these materials. These corresponding $ZT$ values are shown in Table S1 in Supplementary Materials.

The band gaps of the 308 compounds are within the energy region of (0.02 - 5.90 eV). The semiconducting behavior is essential for thermoelectric materials, but there is no a clear line to separate insulators and semiconductors using the band gap. From the previously investigated binary chalcogenides, such as PbX (X=S, Se and Te)[3,7,16,17,22], $Y_2X_3$ (Y=As, Sb and Bi)[78-80], (Sn, Ge)X[13,28,62], $MX_2$ (M=Sn, Zr, Mo, W and Pt)[23,27,81-83], $MX_3$ (M = Ti, Hf and Zr)[24-26,84], etc., (Fig. 1) we could roughly classify semiconductors and insulators. From Fig. 1, we find that these promising thermoelectric materials (the blue and red circles) have a band gap less than 2.5 eV. Thus, 2.5 eV is used as a criterion (the black dashed line in Fig. 1) in our work to separate semiconductors and insulators, and 243 binary chalcogenides are considered as semiconductors. For those binary chalcogenides, we notice that only a small number of the compounds (~20%) have been studied for the thermoelectric performance and an enormous number of materials have not been considered yet (Fig. 1). To find new promising thermoelectric materials, their thermoelectric properties (electrical transport properties and thermal conductivities) should be fully investigated using HT computations. Our defined parameter ($\chi$ and $\gamma$) will be used as effective descriptors to screen compounds with high power factors and strong anharmonicity, respectively.

## IV. Maximum power factor ($PF_{max}$) described by $\chi$

Electrical transport properties ($S$, $\sigma$ and $PF$) are the essential thermoelectric parameters to determine high performance thermoelectric materials. As we mentioned before, the calculations of $\sigma$ and the corresponding $PF$ are limited by $\tau$. By solving BTE under CRTA (BTE-CRTA), Gibbs et al.[39] calculated the carrier effective masses ($m_d^{BT^*}$ and $m_c^{BT^*}$) and the Fermi surface complexity factor ($N_V K^*$) (Eq. 12-14), and they implied a nearly linear relationship between $N_V K^*$ and the BTE-CRTA calculated maximum power factor ($PF_{max}$) at 600K for ~2300 cubic compounds. It seems that their determined $N_V K^*$ can be used as a descriptor to characterize the electrical properties. However, the $N_V K^*$ descriptor do not include the information of the carrier relaxation time. As mentioned above, beyond the CRTA, we can establish a thermoelectric parameter ($\chi$) (Eq. 11) to evaluate $PF_{max}$. We need to first validate the relationship between $\chi$ and $PF_{\max}$.

From Eq. 11, $\chi$ depends on the effective masses ($m_d^*$ and $m_c^*$), which can be either calculated using the BTE calculated electrical properties ($S$ and $\sigma/\tau$ in Eq. 12 and 13) or applying the RB approximation (Eq. 16 and 17). Based on the RB approximation, we can feasibly calculate the DOS effective mass ($m_d^{RB^*}$) and conductivity effective mass ($m_c^{RB^*}$) by fitting the DOS (Eq. 16) and band structures (Eq. 17) within a specific energy range $\Delta E$ at VBM and CBM for *p-type* and *n-type*, respectively. Although the RB calculated effective masses can be easily used in the HT computations, the accuracy should be checked using the BTE calculated results.

In this work, by solving the electron BTE, we calculate $S$ and $\sigma/\tau$ of 243 binary semiconducting chalcogenides at different electron/hole concentrations ($10^{19}$ - $10^{21}$ cm$^{-3}$) and different temperatures (200 – 1000 K). According to Eq. 12, the BTE determined conductivity effective mass ($m_c^{BT^*}$) is calculated, and the corresponding carrier relaxation time ($\tau^{BT}$) is evaluated by Eq. 6. We then can evaluate the electrical conductivity ($\sigma = \frac{\sigma}{\tau}\tau^{BT}$) and power factor ($PF = S^2\sigma$) of each compound. Subsequently, the maximum *p-type* and *n-type* power factors ($PF_{max}$), the optimum carrier concentration ($n_{PF_{max}}$) and the corresponding $m_d^{BT^*}$ and $m_c^{BT^*}$ at different temperatures can be obtain. To calculate $m_d^{RB^*}$ and $m_c^{RB^*}$, the optimum chemical potential ($\mu_{max}$) can be evaluated using the target carrier concentration ($n_{max}$) (Eq. 4), and it is reasonable to choose the fitting energy region as $\Delta E = \mu_{max}$.

Fig. 2 shows the case of the RB calculated *p-type* and *n-type* $m_d^{RB^*}$ and $m_c^{RB^*}$ versus the BTE calculated *p-type* and *n-type* $m_d^{BT^*}$ and $m_c^{BT^*}$ at $n_{max}$ and different temperatures (300 K, 600 K and 900 K). From Fig. 2, for the DOS effective mass ($m_d^*$), the RB calculated $m_d^{RB^*}$ are in excellent agreement with BTE calculated $m_d^{BT^*}$ at different temperatures (Fig. 2a). For the conductivity effective mass ($m_c^*$), even though the data distribution of $m_c^{RB^*}$ versus $m_c^{BT^*}$ is kind of smearing with increasing temperature (Fig. 2b), the agreement is still reasonable. The reason behind the data smearing is that the temperature effect (such as the Fermi-Dirac distribution and the ambipolar electric field effect) is not fully considered in the RB approximation. Therefore, using the RB calculated effective masses ($m_d^{RB^*}$ and $m_c^{RB^*}$) to calculate $\chi^{RB}$ is reasonable.

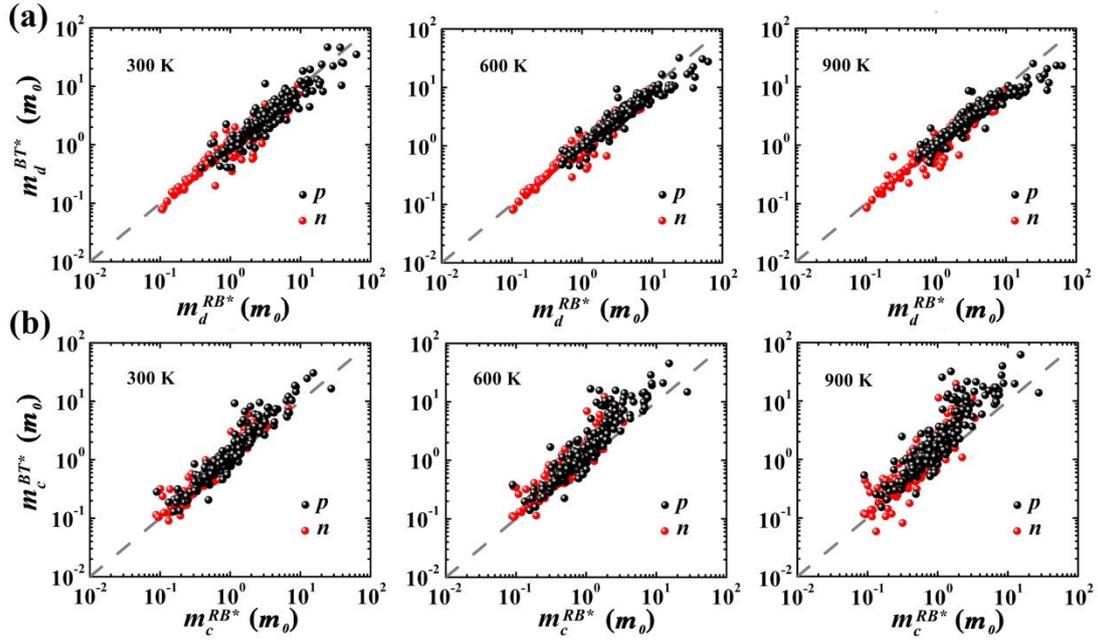

Figure 2. The RB approximation calculated *p-type* and *n-type* $m_d^{RB*}$ and $m_c^{RB*}$ of 243 binary semiconductor chalcogenides versus the corresponding BTE calculated *p-type* and *n-type* $m_d^{BT*}$ and $m_c^{BT*}$ at 300 K, 600 K and 900 K, respectively. $m_0$ is a free electron mass, the black and red dots represent *p-type* and *n-type*, respectively. The gray lines are the ideal case of $m_d^{RB*} = m_d^{BT*}$ and $m_c^{RB*} = m_c^{BT*}$. These data are shown in Supplementary Tables.

Using the RB calculated $m_d^{RB*}$ and $m_c^{RB*}$, we can evaluate $\chi^{RB}$. Fig. 3 shows the data between the RB determined electrical descriptor ($\chi^{RB}$) and the corresponding BTE calculated maximum power factor ($PF_{max}$) for 243 binary semiconductor chalcogenides at 300K, 600K and 900K, respectively. From Fig. 3, we can see a reasonable linear relationship between $\chi^{RB}$ and $PF_{max}$, especially at low temperature. To investigate the possible reason of the error between $\chi^{RB}$ and $PF_{max}$, we also evaluate $\chi^{BT}$ using the BTE calculated $m_d^{BT*}$ and $m_c^{BT*}$ [$\chi^{BT} = \frac{\left(m_d^{BT*}\right)^{3/2}}{\left(m_c^{BT*}\right)^{5/2}} \frac{\hbar k_B^2 \rho v_l^2}{E_d^2}$] and show the relationship between $\chi^{BT}$ and $PF_{max}$, as shown in Fig. S2 in Supplementary Materials. Since $\chi^{BT}$ to $PF_{max}$ are evaluated under the same framework, it shows a very good linear relationships between $\chi^{BT}$ to $PF_{max}$ at different temperatures. This means that the error between $\chi^{RB}$ and $PF_{max}$ is inherited from the error between $m_{d/c}^{RB*}$ and $m_{d/c}^{BT*}$ rather than the definition of $\chi$.

In Fig.3, at higher temperature, the dispersion of those data is becoming larger, which is due to that the temperature related effect is not fully considered in the $m_c^{RB*}$ calculations. Nevertheless, a linear fitting is carried out based on the expression,

$$\log_{10}(PF_{\max}) = k\log_{10}(\chi^{RB}) + c \tag{19}$$

The fitting parameters of $(k, c)$ are (1.06, 2.10), (1.21, 2.11) and (1.28, 2.01) at 300 K, 600 K and 900 K, respectively. Due to the fitting parameters ($k$ and $c$) change small at different temperatures, we can choose $(k, c)$ to be equal (1.18, 2.06) and get a relationship between $\chi^{RB}$ and $PF_{\max}$ as

$$PF_{\max} = 115(\chi^{RB})^{1.18} \tag{20}$$

And the error ($\Delta e = \frac{1}{243}\sum \frac{|PF_{\max}(\chi^{RB}) - PF_{\max}|}{PF_{\max}}$) in predicting the maximum power factor $[PF_{\max}(\chi^{RB})]$ from the descriptor ($\chi^{RB}$) is 1.34 at 300 K. The electrical descriptor ($\chi^{RB}$) can be used as a reasonable descriptor to evaluate the electrical transport properties ($PF_{\max}$).

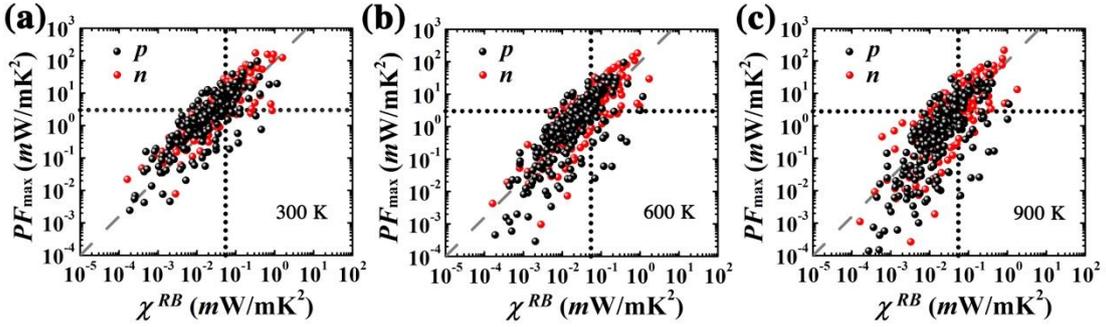

Figure 3. The RB determined descriptor ($\chi^{RB}$) versus the corresponding BTE calculated maximum power factor ($PF_{max}$) for 243 binary semiconductor chalcogenides at 300 K, 600 K and 900 K, respectively. The dashed line represents the fitting relationship between $\chi^{RB}$ and $PF_{max}$. The horizontal and vertical dot lines represent that $PF_{max}$ is 3 $m$W/mK² and $\chi^{RB}$ is 0.05 $m$W/mK², respectively. The black and red dots represent *p-type* and *n-type*, respectively. These data are shown in Supplementary Tables.

## V. Anharmonicity described by $\gamma$

Promising thermoelectric materials require not only high electrical transport properties (such as $PF$) but also low lattice thermal conductivities ($\kappa_l$), which can maintain the temperature gradient across a material. $\kappa_l$ strongly depends on the phonon scatterings or the lattice anharmonicity. Even though the Grüneisen parameter ($\gamma$) is a well-known parameter to characterize the lattice anharamonicity, the accurate $\gamma$ evaluation requires time-consuming qusi-harmonic phonon calculations[67] or solving the second- and third-order force constants[68,69], which is not suitable for the HT calculations to screen the promising thermoelectric material candidates with intrinsically ultralow lattice thermal conductivity. To effectively evaluate the lattice anharmonicity of a compound, we adopt our recently developed elastic-property-estimated Grüneisen approach[51].

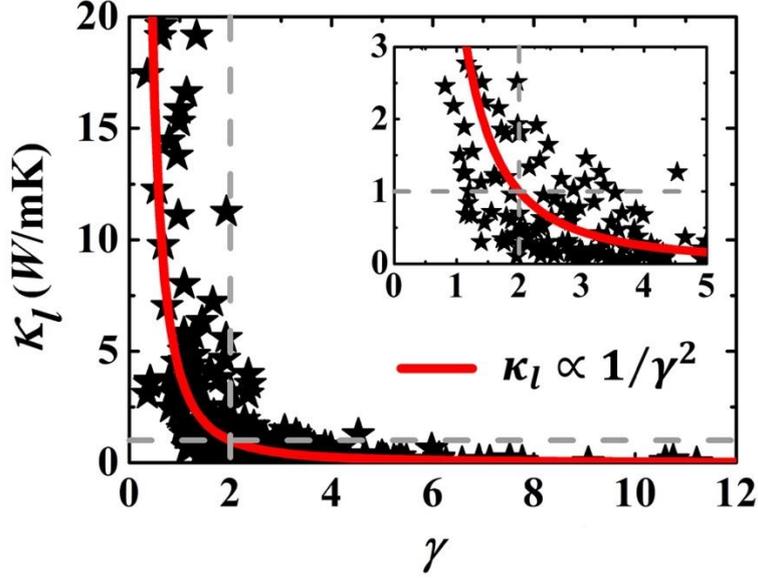

Figure 4. Elastic-property-estimated Grüneisen parameter ($\gamma$) versus the corresponding lattice thermal conductivity ($\kappa_l$) of 243 binary semiconductor chalcogenides at 300K. The grey horizon and vertical dashed lines represent $\kappa_l = 1$ W/mK and $\gamma = 2$. The red curve represents the relationship of $\kappa_l \propto 1/\gamma^2$.

For the 243 binary semiconductor chalcogenides, we calculate their $\gamma$ based on the variation of the elastic properties (B and G) with the volume (Eq. 18). Another thermal parameter, the Debye temperature ($\Theta$), is also required to evaluate the lattice thermal conductivity, and it was calculated using the Anderson approach[57]

$$\Theta = \left[\frac{3m}{4\pi}\right]^{\frac{1}{3}} \frac{h}{k_B} v_a n_a^{-1/3}, \ v_a = \left[\frac{1}{3}\left(\frac{1}{v_l^3} + \frac{2}{v_s^3}\right)\right]^{-\frac{1}{3}}, \ v_l = \sqrt{\frac{B+\frac{2}{3}G}{\rho}}, \ v_s = \sqrt{\frac{G}{\rho}} \quad (21)$$

where $h$ is the Planck constant, $m$ is the number of atoms per volume, and $v_l$, $v_s$ and $v_a$ are the longitude, shear and averaged sound velocity, respectively. Then we insert the elastic-property-estimated $\gamma$ and $\Theta$ into the Slack model[85]

$$\kappa_l = A \frac{m_a \phi n_a^{1/3} \Theta^3}{\gamma^2 T}, \ A = \frac{2.43\times10^{-6}}{1-\frac{0.514}{\gamma}+\frac{0.228}{\gamma^2}} \quad (22)$$

(where $m_a$ is the average atomic mass, $\phi^3$ is the volume of per atom and $n_a$ is the number of atoms in the primitive cell) and evaluate $\kappa_l$. We plot the theoretically calculated $\kappa_l$ versus $\gamma$ in Fig. 4. We notice that even through $\kappa_l$ is a function of many parameters (such as $\Theta$, $m_a$ and $n_a$), the relationship of $(\gamma, \kappa_l)$ clearly shows the well-known $\kappa_l \propto 1/\gamma^2$ behavior. This indicates that $\gamma$ is the dominated parameter in the thermal conductivity evaluation. It is reasonable to use $\gamma$ to simply evaluate $\kappa_l$.

Additionally, it is worth noting that despite the van der Waals (vdW) interaction might play an important role in the calculation of the elastic properties for some layered structures[46], the vdW corrections should has a small effect on $\gamma$, since $\gamma$ is decided by the change of the elastic properties. For SnSe$_2$, our elastic-property-estimated $\gamma$ (2.62) by GGA is comparable with the DFT-D2 vdW method calculated $\gamma$ (1.55 in plane and 2.04 in out-of-plane)[96] and the experimentally measured results (1.13 in plane and 2.36 in out-of-plane)[97]. Thus, our calculated $\kappa_l$ (1.32 W/mK at 300 K) by GGA is comparable with the experimental $\kappa_l$ (~3.1 W/mK in plane and ~1.0 W/mK in out-of-plane at 300 K)[97]. This further means that it is reasonable to use $\gamma$ to describe $\kappa_l$. Moreover, it also needs to point out that since the lattice thermal conductivity methodology using the elastic properties only consider the contributions of acoustic phonons, not the optical phonons[10,11], the evaluated results might lower than the experimental measurements.

## VI. Promising binary chalcogenide thermoelectric materials

The high power factor and low lattice thermal conductivity are the necessary prerequisites for the high $ZT$ value. We have established the power factor descriptor ($\chi$) and lattice thermal conductivity descriptor ($\gamma$). Thus, to determine the possible promising thermoelectric materials through HT computations, we should setup thresholds for the lower limit of $\gamma$ and $\chi$. For a promising thermoelectric material, the value of $ZT$ is expected $ZT > 1$. For $\kappa_l$, 1.0 W/mK (at 300 K, the grey horizon dashed line Fig. 4) has been frequently chosen as the upper limit for a low thermal conductivity for an inorganic crystalline semiconductor[86]. According to the relationship between $\kappa_l$ and $\gamma$ in Fig. 4, we could suggest the lower limit of $\gamma$ should be around 2 (e.g. $\gamma \geq 2$, the grey vertical dashed line in Fig. 4). Therefore, inserting $ZT > 1$ and $\kappa_l \leq 1.0$ W/mK into $ZT = PF/\kappa_l$, we evaluate that the lower limit of the $PF$ is $PF \geq 3$ $m$W/mK$^2$ (the black horizontal dot line in Fig. 3). Based on the fitting relationship between the electrical descriptor ($\chi^{RB}$) and the maximum power factor ($PF_{max}$) (Eq. 20), we estimate that $\chi^{RB}$ should be bigger than 0.05 $m$W/mK$^2$.

After establishing the lower limits descriptors of $\gamma$ and $\chi^{RB}$, at 300 K, we can screen promising thermoelectric materials from the 243 binary semiconductor chalcogenides by plotting $\gamma$ and $\chi^{RB}$ (Fig. 5). Among them, 50 (24 *p-type* and 43 *n-type*) binary chalcogenides are identified as promising thermoelectric materials. The corresponding $\gamma$ and $\chi^{RB}$ of the 50 predicted binary chalcogenides are listed in Supplementary Tables.

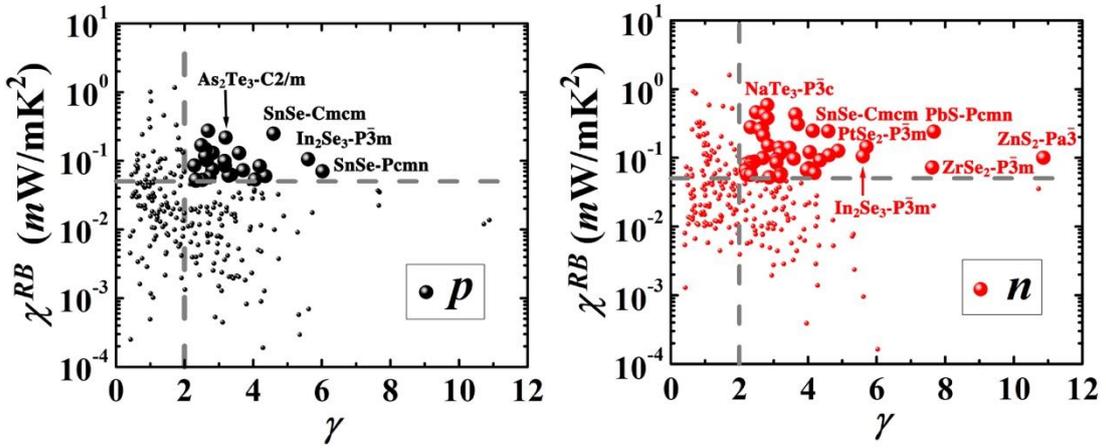

Figure 5. Screening high performance thermoelectric materials within the 243 binary semiconductor chalcogenides using the elastic-property-estimated Grüneisen parameter ($\gamma$) and the RB determined power factor descriptor ($\chi^{RB}$). The black and red spheres represent the theoretically predicted promising compounds for *p-type* and *n-type*, respectively.

To verify the validity of these two descriptors ($\gamma$ and $\chi^{RB}$), we screen promising thermoelectric materials directly using $\kappa_l$ and $PF_{max}$ as well (Fig. S4 in Supplementary Materials). Based on the judgement criteria, $\kappa_l \leq 1$ W/mK and $PF \geq 3$ $m$W/mK$^2$, we predict 47 (24 *p-type* and 43 *n-type*) promising thermoelectric binary chalcogenides. The corresponding $\kappa_l$ and $PF_{max}$ of the 47 predicted binary chalcogenides are listed in Supplementary Tables. Among them, most promising thermoelectric materials (~60%) are successfully included in the simple descriptors ($\gamma$ and $\chi^{RB}$) predictions, as shown in Fig. S3 in Supplementary Materials. This means that the using the simply descriptors ($\gamma$ and $\chi^{RB}$) to simply screen the promising thermoelectric materials in HT computations is reasonable.

Additionally, since the simple CRTA [assuming the carrier relaxation time of different compounds to be a constant (10$^{-14}$ s) at different temperatures and carrier concentrations] was used in the previously thermoelectric calculations[37,39], we need to check its prediction ability. Using the calculated the maximum power factor under the simple CRTA ($PF_{\max}^{CRTA}$, which are listed in Supplementary Tables), we do predict 42 (28 *p-type* and 15 *n-type*) promising binary chalcogenides with $PF_{\max}^{CRTA} \geq 3$ $mW/mK^2$ at 300 K. However, among them, only 11 (5 *p-type* and 6 *n-type*) compounds could be verified by the $\kappa_l$ and $PF_{\max}$ prediction results. The well-known thermoelectric materials (such SnSe, GeSe and $Bi_2Te_3$) cannot be predicted by the CRTA method. And some promising compounds predicted by $PF_{\max}^{CRTA}$ have large lattice thermal conductivities (>50 W/mK). This means that the evaluations of $\tau$ and $\kappa_l$ have a significant improvement on the prediction ability.

For the theoretically predicted 50 promising thermoelectric materials using the descriptors ($\chi^{RB}$ and $\gamma$), we successfully predict 34 previously experimentally and theoretically investigated thermoelectric materials, such as $Bi_2(S, Se, Te)_3$[78], $Sb_2Te_3$[79], $As_2(Se, Te)_3$[80], $In_2Se_3$[87], $Ga_2Te_5$[88], Ga(S, Se)[89], $ReSe_2$[90], (Ge, Sn)(S, Se)[13,28,62], M(S, Se, Te)$_2$ (M= Hf, Mo, W, Pt and Zr)[81-83], Sn(S, Se)$_2$[23,91], and $ZrSe_3$[84], etc. The thermoelectric properties ($\kappa_l$ and $PF_{\max}$, $\gamma$ and $\chi^{RB}$) of the 50 promising thermoelectric materials and the corresponding references are listed in Table S2 in Supplementary Materials. In addition, we notice that the famous PbTe (225) compound is not predicted in the current work, but in some previous HT works[39,45]. This is because that the previous works mainly consider electrical properties[39] or using the large thermoelectric parameter ($\beta$) with an optimal $\kappa_l$ range (~1–10 W/mK) to screen promising thermoelectric materials. In this work, if we only considered the electrical properties, the PbTe compound can be screened as a promising thermoelectric material as well, due to its high $PF_{\max}$ and large $\chi^{RB}$ [$PF_{\max}$ = 43.0 and 10.9 $mW/mK^2$, $\chi^{RB}$ = 0.137 and 0.053 $mW/mK^2$ at 300 K for *p-type* and *n-type*, respectively], indicating a good electrical transport property. This is consistent with the previous works. However, our screening criteria are stricter: especially in the lattice thermal conductivity ($\kappa_l$ <1 W/mK) or strong anharmonicity ($\gamma$ >2). Following our strict criteria, PbTe cannot be screened due to its small Grüneisen parameter ($\gamma$ =1.13) and high thermal conductivity (>1 W/mK) which is in good agreement with the experimentally measured ~2.3 W/mK at 300 K[92]. Similarly, other some previously reported thermoelectric materials [PbS (225) and PbSe (225)] with small Grüneisen parameter ($\gamma$ < 2) are not predicted. Additionally, due to the prediction errors between the descriptors ($\gamma$, $\chi^{RB}$) and ($\kappa_l$, $PF_{\max}$), some previously reported thermoelectric materials [such as, InSe, In$_4$Se$_3$, and Bi$_2$Se$_3$ (R-3m)] cannot be predicted using $\gamma$ and $\chi^{RB}$, but can be predicted using $\kappa_l$ and $PF_{\max}$ (as listed in Supplementary Tables).

Our HT computations provide an alternative way to screen promising thermoelectric materials with both the ultralow lattice thermal conductivity (or large Grüneisen parameter) and high power factor (or large $\chi^{RB}$). Except those previously experimentally and theoretically investigated compounds, some previously unreported binary chalcogenides [such as, Ba$_2$S$_3$, BeSe, Bi(S, Se)$_2$, Cs$_2$Te$_4$, K$_2$Te$_3$, NaTe$_3$, Sb$_2$Te$_2$, Sr(S, Te), Te$_{16}$Si$_{38}$, TlS, ZnSe, and Zn(S, Se)$_2$ etc.] have been predicted as promising thermoelectric materials (9 and 14 compounds for *p-type* and *n-type*, respectively). Among them, due to the simple structure and cheap price of ZnSe$_2$, we have carefully investigated its thermoelectric properties by solving the full electron and phonon BTEs and find that ZnSe$_2$ dose have exhibit excellent thermoelectric properties.[93] This means that we predicted new thermoelectric materials can be validated by the precise method.

There are two factors leading to our predicted new compounds. (1) In the previous HT computations[37-39], only using electrical properties to screen the promising thermoelectric materials, could result in ambiguous predictions. Our screening methodologies consider all important thermoelectric properties (the electrical transport and the thermal properties) and the prediction results are more reasonable. (2) Most of the previous HT works have mainly focused on ternary compounds[37,42], half-Heuslers[38], sintered materials[94], transition metal silicides[95], A$_1$B$_1$[45], and quasi-2D compounds[46], etc. Among them, only a small part of binary chalcogenides were considered. In other words, considering a much larger binary chalcogenide data base and more reasonable screening method in our HT computations, we could predict distinct high performance thermoelectric materials. Nevertheless, future work is worthwhile to discuss the dopability of those predicted promising thermoelectric materials.

## VII. Possible reasons for the promising materials

For our theoretically predicted compounds, In$_2$Se$_3$-P$\bar{3}$m1, MSe$_2$-P$\bar{3}$m1 (M=Pt and Zr), PbS-Pcmn and ZnS$_2$-Pa$\bar{3}$ have strong anharmonicity (the large Grüneisen parameters $\gamma > 5$, and their corresponding $\gamma$ are shown in Fig. 5 and Fig. S3), indicting ultralow lattice thermal conductivity. The strong anharmonicity of In$_2$Se$_3$-P$\bar{3}$m1 and MSe$_2$-P$\bar{3}$m1 are due to their layered structures (their crystal structures are list in Fig. S5 in Supplementary Materials), serving as stress buffer to strongly scatter phonons as discussed in SnSe[13]. Additionally, compounds with soft lattices (long bond lengths and high coordination numbers) are suggested to exhibit strong anharmonicity[12]. Thus, the strong anharmonicity of ZnS$_2$-Pa$\bar{3}$ may be due to the complex geometry structure (Zn has six neighboring S) and the soft bonding interaction between Zn and S atom (the Zn-S bond length is longer than those in ZnS-F$\bar{4}$3m and ZnS-P63mc, as shown in Fig. S6 in Supplementary Materials). For the strong anharmonicity of PbS-Pcmn may be due to the soft bonding interaction between Pb and S atom (the Pb-S bond lengths are longer than those in PbS-Fm$\bar{3}$m, as shown in Fig. S7 in Supplementary Materials). In addition, for electrical transport properties, compounds with large $\chi$ are considered exhibiting high power factors. In this work, GeSe-Pcmn, ZrSe$_3$-P121m1, As$_2$Te$_3$-C12/m, NaTe$_3$-P$\bar{3}$c1, PbS-Pcmn and SnSe-Cmcm have high maximum power factors ($PF_{max} > 30.0$ $m$W/mK$^2$). Through analyzing their band structures (as shown in Fig. S8 in Supplementary Materials), we can find that their big $N_V K^*$ or small $m_c^*$ can contribute their large $\chi$, due to $\chi = \frac{(m_d^*)^{3/2}}{(m_c^*)^{5/2}} \frac{\hbar k_B^2 \rho v_l^2}{E_d^2} \propto \frac{N_V K^*}{m_c^*}$ (their $\chi^{RB}$ are shown in Fig. 5 and Fig. S3). Further experimental measurements should be carried on for these predicted promising thermoelectric materials.

Moreover, to establish a simple intuition of screening high performance thermoelectric materials, we statistically analyze all 243 binary semiconductor chalcogenides (Y$_n$X$_m$, X=S, Se, Te) based on their space group numbers (*N*) and stoichiometries (n:m), as shown in Fig. 6. We find the high performance thermoelectric materials are mainly grouped in the low symmetry regions [monoclinic (*N*=~12) and orthorhombic (*N*=~36, 62) crystals] and the high symmetry regions [trigonal (*N*=~164), hexagonal (*N*=~194) and cubic (*N*=221) crystals]. It is interesting to notice that even through the tetragonal region is the largest one in the plot, compounds from this crystal do not prefer for high performance thermoelectric. For the stoichiometry (n:m), compounds with n:m = 0.5-1.0 exhibit excellent thermoelectric properties. This means that the crystal type and stoichiometry can provide a useful preliminary judgement for the thermoelectric performance of a compound.

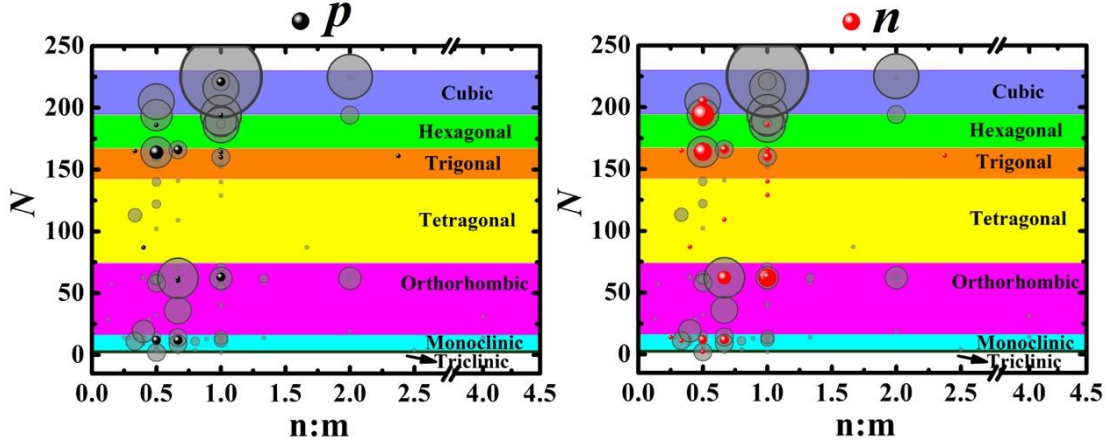

Figure 6. Statistically analyzing of all 243 binary semiconductor chalcogenides ($Y_nX_m$, X=S, Se, Te) based on their space group numbers ($N$) and stoichiometries (n:m). The grey, black and red circles represent all the binary chalcogenides, the HT predicted *p-type* and *n-type* thermoelectric materials, respectively. The size of circle represents the number of compounds at ($N$, n:m). Additionally, the dark grey (its region is too narrow), cyan, pink, yellow, orange, green and blue regions represent the seven crystal systems (triclinic, monoclinic, orthorhombic, tetragonal, trigonal, hexagonal and cubic), respectively.

## VIII. Conclusion

Utilizing high-throughput (HT) computations, we have investigated the thermoelectric properties of 668 binary chalcogenides. In order to improve the prediction ability in thermoelectric HT computations, we include the electronic relaxation time using the deformation potential method and evaluate the Grüneisen parameter (or lattice anharmonicity) by computing the elastic properties. The important effective masses are calculated based on the rigid band approximation, and the calculated results are in good agreement with those calculated using the electron Boltzmann transport theory. We establish a feasible electrical descriptor ($\chi$) to characterize the maximum thermoelectric power factor, and use the anharmonicity descriptor ($\gamma$) to predict the low thermal conductivity. Applying the two computational efficient descriptors ($\chi$ and $\gamma$) in 243 binary semiconductor chalcogenides, we screen 50 promising thermoelectric materials with both excellent electronic properties and intrinsically ultralow thermal conductivities. From these theoretically determined compounds, we successfully predict not only previously experimentally and theoretically investigated promising thermoelectric materials but also 9 and 14 previously unreported binary chalcogenides as promising *p-type* and *n-type* thermoelectric materials, respectively. These new thermoelectric candidates deserve further investigations experimentally and theoretically. Moreover, our work provides reliable thermoelectric descriptors ($\chi$ and $\gamma$) for screening high performance thermoelectric materials through HT computations in the future.

## Acknowledgments

This work was supported by the National Natural Science Foundation of China, Grant No. 11774347 and 11474283. The calculations were performed in at the Center for Computational Science of CASHIPS, the ScGrid of Supercomputing Center and Computer Network Information Center of the Chinese Academy of Science and the Tianhe-2JK at the Beijing Computational Science Research Center. The authors acknowledge with gratitude the assistances and guidance given by Prof. Georg K. H. Madsen.